\documentclass[twoside]{article}
\usepackage{fleqn,espcrc2}
\usepackage{graphicx}
\usepackage{here}

\newcommand{\AmS}{{\protect\the\textfont2
    A\kern-.1667em\lower.5ex\hbox{M}\kern-.125emS}}										
\def\beq{\begin{equation}}
\def\eeq{\end{equation}}
\def\bea{\begin{eqnarray}}
\def\eea{\end{eqnarray}}
\def\bq{\begin{quote}}
\def\eq{\end{quote}}

\def\nnb{\nonumber}
\def\ga{\left(}
\def\dr{\right)}

\def\rar{\rightarrow}
\def\lrar{\Longrightarrow}
																
\def\nnb{\nonumber}
\def\la{\langle}
\def\ra{\rangle}
\def\nin{\noindent}
\def\ba{\vspace*{-0.2cm}\begin{array}}
\def\ea{\end{array}\vspace*{-0.2cm}}

\def\b{$\bullet~$}
\def\als{\alpha_s}

\def\gg2{ \la\alpha_s G^2 \ra}
\def\gg3{g^3f_{abc}\la G^aG^bG^c \ra}
\def\ggg4{\la\als^2G^4\ra}

\title
{\bf{\boldmath
{\Large Power corrections to $\alpha_s(M_\tau),~\vert V_{us}\vert$ and $\bar m_s$   
} }}

\author
{ 
Stephan Narison\thanks{Email: snarison@yahoo.fr}  \address {\footnotesize Laboratoire de Physique Th\'eorique et Astroparticules, CNRS - IN2P3 \& Universit\'e
de Montpellier II, Case 070, Place Eug\`ene Bataillon, 34095 - Montpellier Cedex 05, France},
}

\begin{document}

\pagestyle{myheadings}
\markright{ }
\begin{abstract}
\noindent
We re-examine recent determinations of power corrections from $\tau$-decay and confront the results with the existing ones from QCD spectral sum rules (QSSR). We conclude that contrary to the QSSR analysis, which lead to $\la \alpha_s G^2\ra=(6.8\pm 1.3)10^{-2}$ GeV$^4$, $\tau$-decay is not a good place for extracting the gluon condensate
due to its extra $\alpha_s^2$ coefficient  which suppresses its contribution in  this process.
Results from $e^+e^-$ sum rules and $\tau$-decay:  $\rho\alpha_s\la\bar uu\ra^2= (4.5\pm 0.3) ~10^{-4}~{\rm GeV}^6$, where $\rho=3.0\pm 0.2$ confirm the deviation from the vacuum saturation estimate of the four-quark condensate.
 ``Non-standard" power corrections
 (direct instantons, duality violation and tachyonic gluon mass) beyond the SVZ-expansion,
partially cancel out  in the V+A hadronic $\tau$-decay channel, which gives at order $\als^4$: $\alpha_s(M_\tau)=0.3249~(29)_{\rm ex}(75)_{\rm th} $  leading to $\als(M_Z)\vert_\tau=0.1192~(4)_{\rm ex}(9)_{\rm th}$,
in remarkable agreement with (but more accurate than) $\alpha_s(M_Z)\vert_Z=0.1191~(27) $ obtained at the same $\alpha_s^4$  order from the $Z$-width and the global fit of electroweak data. 
Finally, the r\^ole of the tachyonic gluon mass in the determinations of $\vert V_{us}\vert$ from $\tau$-decay and of $\bar m_s$  from $\tau$-decay, $e^+e^-$ and (pseudo)scalar channels is  emphasized. 
\end{abstract}
\maketitle
\section{Introduction}
\nin
Hadronic $V+A$ $\tau$-decay is expected to provide the most accurate determination of $\alpha_s$ once its value obtained at the $\tau$-mass is runned until the $Z$-mass
\cite{BNP0,BNP,LEDIB,PIVOV,PICH}.
The extraction of $\alpha_s$, at this relatively low scale $M_\tau$, can become feasible, as theoretically, the $V+A$ channel is less sensitive to the QCD non-perturbative effects than the individual $V$ and $A$ channel (if one uses the OPE \`a la SVZ \cite{SVZ}) and to the values of the quark and gluon condensates obtained from an overall fit of the hadronic channels \cite{SVZ,SNB}, $e^+e^-$ \cite{SNe,TARRACH,PEROTTET,BORDES,MENES} and $\tau$-decay \cite{DOMI,ALEPH,OPAL,DAVIER,SNZ1,RAF04,SNVA} data, light baryon sum rules \cite{DOSCH}, heavy quarkonia sum rules \cite{SVZ,BELL,YND,IOFFE} and heavy-quarkonia mass-splittings \cite{SNG}.  In addition, the dominant theoretical uncertainties on the $\alpha_s$ determination, though relatively modest at the $\tau$-mass (10\%), becomes tiny at the $Z$-pole as it decreases as $1/\log^2$. 
More quantitatively, the non-strange $\Delta S=0$ component of the $\tau$-hadronic width can be expressed as:
\bea
R_{\tau} &\equiv& {\Gamma(\tau\rightarrow\nu_\tau+{\rm hadrons}\vert_{\Delta S=0})}\over {\Gamma(\tau\rightarrow l+\bar\nu_l+\nu_\tau)}\nnb\\
  &=&  3 |V_{ud}|^2 S_{EW}\times\nnb\\
  &&(1+\delta^{(0)} + \delta'_{EW}+ \delta_m^{(2)}+ \delta_{\rm svz}+\delta_{\rm nst})~,
\eea
where  $\delta_m^{(2)}$ is the light quark mass corrections, 
\beq
\delta_{\rm svz}\equiv \sum_{D=4}^8\delta^{(D)}~,
\eeq
is the sum of the non-perturbative (NP) contributions of dimension $D$ within the SVZ expansion \cite{SVZ}, while $\delta_{\rm nst}$ are some eventual NP effects not included into $\delta_{\rm svz}$.
We shall use: $|V_{ud}| =  0.97418  \pm \, 0.00027$ \cite{PDG}.
The electroweak corrections are:
\beq
S_{EW} = 1.0198 \pm 0.0006~ \cite{MARCIANO} ~~{\rm and}~~ \delta'_{EW} = 0.001 ~\cite{LI}. 
\eeq
$\delta^{(0)}$ is the perturbative correction, while
$\delta_m$ is the light quark masses corrections.
They will be discussed in details in the next paragraph.
\section{SVZ power corrections in hadronic $\tau$-decay}
\nin
Using the QCD expressions compiled in BNP \cite{BNP}, the light quark masses corrections read:
\bea
M_\tau^2\delta^{(2)}_{m,V/A}&=&-8\ga 1+{16\over 3}a_s\dr \ga \bar m_u^2+\bar m_d^2\dr\nnb\\
&&\pm 4\ga 1+{25\over 3}a_s\dr  \bar m_u\bar m_d~,
\eea
where $\bar m_q$ and $a_s\equiv (\alpha_s/\pi)$ are respectively the light quark running masses and QCD running coupling evaluated at $M_\tau$. Using the recent determinations of the running mass in units of MeV and evaluated at 2 GeV \cite{SNQ,SNmse,SNmstau,SNQ2}:
\beq
\bar m_s= 96.1\pm 4.8~,~~\bar m_d= 5.1\pm 4~,~~\bar m_u= 2.8\pm 0.2~,
\label{eq:msaverage}
\eeq
one can deduce the small and negligible corrections in $10^{-4}$:
\beq
\delta^{(2)}_{m,V}= -(3.15\pm 0.51)~,~~~~\delta^{(2)}_{m,V+A}= -(4.22\pm 0.68)~.
\label{eq:mass}
\eeq
A theoretical estimate of the quark masses contributions has been often used as input in existing analysis of $\tau$-decay data \cite{ALEPH,OPAL,DAVIER}.
The non-perturbative corrections \` a la SVZ \cite{SVZ} read to leading order in $m_q$ and $\alpha_s$ \cite{BNP}\footnote{We shall use the complete QCD expression in BNP \cite{BNP} for the numerics.}:
\bea
 M_\tau^4\delta^{(4)}_{V/A}&=&{11\over 4} \pi a_s^2{\la \alpha_s G^2\ra}\nnb\\
 &&+ 16\pi^2{ (m_u \mp m_d) \la\bar \psi_u\psi_u\mp \bar \psi_d\psi_d\ra } \nnb\\
M_\tau^6\delta^{(6)}_{V/A}&=& \ga \begin{array}{ c}7\\ -11\end{array}  \dr{256\pi^3\over 27}\rho{\alpha_s\la \bar \psi\psi\ra^2}\nnb\\
M^8_\tau\delta^{(8)}_{V}&\approx& M^8_\tau\delta^{(8)}_{A}\approx -{26\over 162}\pi^2{\la \alpha_s G^2\ra^2}.
\label{eq:np}
\eea
{\scriptsize
\begin{table}[H]
\setlength{\tabcolsep}{0.12pc}
 \caption{\scriptsize SVZ power corrections from $\tau$-decay compared 
with the ones from $e^+e^-$ 
and QSSR analysis of the other hadronic channels. 
FO and CI  correspond to $\alpha_s(M_\tau)=0.331\pm 0.013$ and $0.350\pm 0.010$. 
ALEPH fits come from \cite{ALEPH,DAVIER}, while OPAL fits are from \cite{OPAL}. We take the average 
of different results and take the quadratic mean of the error (bold face) when the different fits are in good agreement, while in the case where some of the results are not significant, we only consider the most accurate fit (boldface).  The final sum $\delta_{\rm svz}$ comes from the average or/and from the most accurate determinations.}

\begin{tabular}{lcccc}
&\\
\hline
\hline
\\
NP$\times 10^3$ &  \multicolumn{2}{c}{ V}& \multicolumn{2}{c}{ V+A} \\
 \\ &$\tau$-decay&$e^+e^-\oplus$QSSR  &$\tau$-decay&$e^+e^-\oplus$QSSR \\
\\
\hline
\hline
\\
$\bf \delta^{(4)}$&&${\bf 0.87\pm 0.09}$&&$-2.01\pm 0.11$ \\
{\bf ALEPH}\\
FO05&$0.68\pm0.10$&&$-2.4\pm 0.1$&\\
CI05&$0.41\pm 0.12$&&$-2.7\pm 0.1$&\\
CI08&$0.01\pm 0.15$&&$-3.0\pm 0.1$&\\
{\bf Average}&& &$\bf (\tau+e^+e^-)$&${\bf -2.53\pm 0.06}$\\
&\\
\hline\\
&\\
$\bf\delta^{(6)}$&&$41.3\pm8.4$&&$-$ \\
{\bf ALEPH}\\
FO05&$27.0\pm 2.5$&&$-1.6\pm 2.5$&\\
CI05&$28.5\pm 2.2$&&$-2.1\pm 2.2$&\\
CI08&$26.8\pm 2.0$&&${ \bf -3.7\pm 1.7}$&\\
{\bf OPAL}&\\
FO99&$27.1\pm 6.6$&&$2.8\pm 8.9$&\\
CI99&$25.6\pm 3.4$&&$1.2\pm 5.6$&\\
{\bf Average}&$\bf (\tau+e^+e^-)$&${\bf 29.4\pm 2.0 }$&&$\bf -6.1\pm 0.9$\\
&\\
\hline\\
&\\
$\bf\delta^{(8)}$&&$-15\pm 6$&&$-$\\
{\bf ALEPH}\\
FO05&$-8.6\pm 0.6$&&$~~0.1\pm 0.5$&\\
CI05&$-9.0\pm 0.5$&& $-0.03\pm 0.05$&\\
CI08&$-8.0\pm 0.5$&& ${\bf ~~0.81\pm 0.36}$&\\
{\bf OPAL}&\\
FO99&$-8.5\pm 1.8$&&$-1.5\pm 3.7$&\\
CI99&$-8.0\pm 1.3$&& $-1.0\pm 3.3$&\\
{\bf Average}&$\bf (\tau+e^+e^-)$&${\bf -9.5\pm 1.1 }$ \\
&\\
\hline
&\\
$\bf\delta_{\rm svz}$&  \\
{\bf Average}&$\bf (\tau+e^+e^-)$&${\bf -21.1\pm 1.9 }$&&${\bf -7.8\pm 1.0}$ \\

&\\
\hline
\hline
\end{tabular}
\label{tab:svz}
\end{table}
}
\nin
where $\la \bar \psi\psi\ra\equiv \la \bar uu\ra\simeq \la \bar dd\ra$ and $\rho \simeq 2-3$ \cite{SNB,SNe,TARRACH,PEROTTET,SNZ1,RAF04,SNVA,DOSCH} indicates a deviation from the vacuum saturation of the four-quark condensate estimate. Vacuum saturation has been assumed for estimating the dimension $D$=8 condensates which is expected to be a very rough approximation \cite{SNe,SNZ1,RAF04,SNVA}.  We compare in Table \ref{tab:svz}, the size of the given dimension non-perturbative (NP) contributions obtained from a fit of the $\tau$-decay data with the ones obtained  from QCD spectral sum rules (QSSR)  \cite{SVZ,SNB} using the I=1 component of the $e^+e^-$ data \cite{SNe}.  Some comments follow:
\\
\b {\bf General remarks} \\
The results from QSSR have been obtained using the ratio of the Laplace/Borel (LSR)sum rules where  radiative corrections tend to cancel out. Therefore, contrary to $\tau$-decay, the result will be only slightly affected by the truncation of the PT series. One can inspect in Table \ref{tab:svz} that there is a fair agreement (within the errors) between the different results from LSR used in $e^+e^-$ and from $\tau$-decay data analysis. One can notice that $\tau$-decay analysis can lead to inaccurate results for some NP effects in the channels where they are
tiny. However, the smallness of these NP effects in  the V+A channel  makes it a best place for extracting $\alpha_s$. \\
\b {\bf Gluon condensate} \\
An attempt to extract this quantity from $\tau$-decay in different channels \cite{ALEPH,OPAL,DAVIER} leads to unreliable results because its value which is expected to be universal differs from one channel to another. In units of $10^{-2}$ GeV$^4$, $\tau$-decay gives in the V, A and V+A channels:
\bea
\la a_s G^2\ra \simeq  && -(0.8\pm 0.4)_V~, ~~-(2.2\pm 0.4)_A~, \nnb\\
&&-(1.5\pm 0.3)_{V+A}~.
\eea
One can also worry on the negative sign which differs from the one obtained from different  QSSR analysis \cite{SVZ,SNB,SNe,TARRACH,PEROTTET,BORDES,MENES,DOMI,BELL,YND,IOFFE,SNG}\footnote{Using arguments based on magnetic confinement, one can argue that $\la \alpha_s G^2\ra$ is positive \cite{NAMBU}. For a review, see e.g. \cite{ZAKA}.}. This  can be due to the fact that its 
contribution in these $\tau$-decay channels is difficult to extract from the data as it acquires an extra $\alpha_s^2$ correction and then becomes relatively tiny compared to the other condensate effects. Indeed, from Eq. (\ref{eq:np}), one can notice that it is one order of magnitude smaller than the $m\la\bar\psi\psi\ra$ quark  and of the four-quark condensate  effects.  Another problem may arise that in ALEPH and OPAL analysis, one has to do simultaneously a fit of many parameters in the $\tau$-decay analysis ($\alpha_s, ~D=4,6$ and 8 condensates). This is not the case of the QSSR analysis of $e^+e^-$ within LSR, which has a stronger sensitivity to $\la \alpha_s G^2\ra$ and then permits its robust estimate. In these QSSR analysis, it has been found from different analysis a strong correlation between the $D=4$ and $D=6$ condensate contributions. Using the average of the ratio of the gluon and four-quark condensates  \cite{SNe} determined in \cite{TARRACH,PEROTTET,BORDES,MENES,DOMI}:
\beq
r_{46}\equiv {\la\alpha_s G^2\ra\over \rho\alpha_s\la\bar uu\ra^2}  =(106\pm 12)~{\rm GeV}^{-2}~,
\label{eq:ratio}
\eeq
which reduces the analysis to a one-parameter fit \footnote{The fit has been performed using a standard Mathematica least-square fit program.}, the LSR analysis of $e^+e^-$ data gives \cite{SNe}:
\beq
\la \alpha_s G^2\ra \vert_{e^+e^-}= (6.1\pm 0.7)10^{-2}~{\rm GeV}^4~,
\eeq
while the heavy quarkonia mass-splittings \cite{SNG} lead to:
\beq
\la \alpha_s G^2\ra \vert_{\rm heavy}= (7.5\pm 2.5)10^{-2}~{\rm GeV}^4~,
\eeq
from which we deduce the average: 
\beq
\la \alpha_s G^2\ra = (6.8\pm 1.3)10^{-2}~{\rm GeV}^4~,
\label{eq:e+e-}
\eeq
where  the errors have been averaged quadratically. 
Though higher by about a factor 1.8 than the original SVZ value :
\beq
\la \alpha_s G^2\ra \vert_{\rm svz}\simeq 3.8\times10^{-2}~{\rm GeV}^4~,
\eeq
the previous estimate is at the border of  the lower bound allowed by Bell \& Bertlmann \cite{BELL} in the analysis of the heavy quark sum rules and is in line with results from Finite Energy Sum Rules (FESR) analysis \cite{PEROTTET} or some other methods \cite{BORDES,MENES} in $e^+e^-$ data  and with  the available lattice calculations\cite{RAKOW}~\footnote{A more exhaustive though still incomplete comparison of the QSSR estimate of $\la \alpha_s G^2\ra$ can be found in Table 2 (page 543) of the Cambridge book in \cite{SNB}. In \cite{IOFFE}, a correlation between the gluon condensate and $m_c$ has been given. This value would correspond to a lighter value of $m_c$ which is still in the range of values given by PDG \cite{PDG} and from the D-meson sum rules \cite{SNFD}. Further analysis to clarify this issue is needed.}:
\beq
\la \alpha_s G^2\ra \vert_{\rm latt}\approx (12.6\pm 3.1)10^{-2}~{\rm GeV}^4~.
\eeq
 Using the value in Eq. (\ref{eq:e+e-}) and the ones of light quark masses and condensates into the QCD expressions of $\delta^{(4)}_{V,V+A}$, one can deduce the QSSR predictions  for $\delta^{(4)}_V$ and $\delta^{(4)}_{V+A}$ given in Table \ref{tab:svz},
which agree fairly with the range of values  from $\tau$-decay \cite{ALEPH,OPAL,DAVIER}. 
 \\
\b {\bf Four-quark condensate}  \\
Using the ratio in Eq. (\ref{eq:ratio}) and the gluon condensate value in Eq. (\ref{eq:e+e-}), one can deduce:
\beq
\rho\alpha_s\la\bar uu\ra^2= (6.4\pm 1.3) ~10^{-4}~{\rm GeV}^6~.
\eeq
Using this value into the theoretical expression of $\delta^{(6)}_V$, one obtains the QSSR predictions in Table \ref{tab:svz} ($e^+e^-$ column), where one can notice a nice agreement between the contributions of the $D=6$ condensates from the LSR in $e^+e^-$ and from $\tau$-decay. Therefore, we can consider as a final result, their average :
\bea
\delta^{(6)}_V&=&(29.4\pm 2.0)10^{-3}\lrar\nnb\\
 \rho\alpha_s\la\bar uu\ra^2&=&(4.5\pm 0.3)10^{-4}~{\rm GeV}^6~.
 \label{eq:4quark}
\eea
Using the value of $\alpha_s$ and $\la\bar uu\ra$ at $M_\tau$, one obtains:
\beq
\rho=3.0 \pm 0.2~,
\label{eq:rhofactor}
\eeq
confirming the deviations of a factor $2\sim 3$ obtained in \cite{SNB,SNe,TARRACH,PEROTTET,BORDES,MENES,SNZ1,RAF04,SNVA,DOSCH} from the vacuum saturation \cite{SVZ} of the value of the four-quark condensate.
In the V+A channel, the existing results quoted in Table \ref{tab:svz} are inaccurate except the most recent one from \cite{DAVIER}. Considering as the final value, its average with the one obtained from Eq. (\ref{eq:4quark}) and the expression in Eq. (\ref{eq:np}), we obtain the value in Table \ref{tab:svz}~:
\beq
\delta^{(6)}_{V+A}=-(6.1\pm 0.9)10^{-3}~.
\eeq
\b {\bf Dimension $D=8$ condensates} \\
There is also a good agreeement in the extraction of the $D=8$ condensates from the vector component of $\tau$-decay and QSSR in $e^+e^-$, which allows us to take the average quoted in Table \ref{tab:svz} ($e^+e^-$ column).  
In the V+A channel, most of the $\tau$-decay results are inaccurate but suggest a value much smaller than the one from the vector channel. As the theoretical relation in Eq. \ref{eq:np} is expected to be (a posteriori) very inaccurate, we shall consider as a value of $\delta^{(8)}_{V+A}$, the most accurate estimate from \cite{DAVIER}:
\beq
\delta^{(8)}_{V+A}=-(0.81\pm 0.36)10^{-3}~.
\eeq
Both results from the V and V+A channels indicate large deviations from the vacuum saturation suggested by Eq. (\ref{eq:np}). Analogous results have been 
obtained from the LSR and FESR analysis of the V-A channel \cite{SNZ1,RAF04,SNVA}. \\
\b {\bf Sum of the SVZ power corrections} \\
Determinations of the NP corrections from the $e^+e^-$ sum rules and $\tau$-decay data agree, in most case, indicating the consistency of the whole picture. However, the failure of $\tau$-decay for extracting $\la \alpha_s G^2\ra$, can be mainly due to the $\alpha_s^2$ suppression of its effect making its extraction more delicate. From Table \ref{tab:svz}, we deduce:
\beq
\delta_{\rm svz}\equiv \sum_{D=4}^8\delta^{(D)}=-(7.8\pm 1.0) 10^{-3}~.
\eeq
This result is comparable with the direct fit from ALEPH and OPAL \cite{ALEPH,OPAL}, where the new result is \cite{DAVIER}:
\beq
\delta_{\rm svz}\equiv \sum_{D=4}^8\delta^{(D)}=-(5.9\pm 1.4) 10^{-3}~.
\eeq
\section{Sum of standard power corrections }

\nin
We add, to these SVZ power corrections, the ones from the pion and $a_0(980)$ poles into the longitudinal part of the spectral function. It can be written as \cite{PRADES} in units of $10^{-3}$:
 \bea
 \delta_{\pi}&=&-16\pi^2{f_\pi^2m_\pi^2\over M^4_\tau}\ga 1-{m_\pi^2\over M^2_\tau}\dr^2
 = -(2.65\pm 0.05)\nnb\\
  \delta_{a0}&=&-(0.02\pm 0.01)~,
 \eea
 where we have used $f_\pi=93.28$ MeV and $f_{a0}=(1.6\pm 0.5)$ MeV \cite{SNB}. Therefore, the total sum of the standard power corrections reads:
 \beq
 \delta_{\rm st}\equiv \delta_{\rm svz}+\delta^{(2)}_{\rm m}+\delta_\pi+\delta_{a0}=-(10.9\pm 1.1)10^{-3}~.
 \label{eq:pcst}
 \eeq
\section{$\als(M_\tau)$ from $\tau$-decay using the SVZ expansion}
 \nin
Here, we shall be concerned with the perturbative correction $\delta^{(0)}$  which can be expressed in terms of
$R_{e^+e^-}$ as:
\beq
1+\delta^{(0)}={2\over M_{\tau}^2}\int_0^{M_{\tau}^2}{dt}
\left(1-\frac{t}{M_{\tau}^2}\right)^2
\left( 1+ \frac{2 t }{M_\tau^2}
\right)
R_{ee}.
\label{equivalent.repr}
\eeq
\nin
 The ratio 
$R_{ee}\equiv \sigma(e^+e^-\to {\rm hadrons}) / \sigma(e^+e^-\to\mu^+\mu^-)$ 
is 
expressed through the absorptive part of the 
correlator of the electromagnetic current $J_\mu$:
\begin{eqnarray}
R_{ee} &=& 12\pi\, \mathrm{Im}\, \Pi(-t- i\epsilon) ~,
\\
3\, Q^2\, \Pi(Q^2) &=& 
i\int d^4 x~e^{iq\cdot x}\langle 0|{\rm  T}J_\mu(x)J^\mu(0)|0\rangle
{},
\end{eqnarray}  
with $Q^2 = -q^2$. 
At  ${\cal O}(\alpha_s^4)$, its  numerical form for $n_f$ flavours is :
\bea
R_{ee} &=& 1  + a_s +  (1.9857 - 0.1152\, n_f)\, a_s^2 
\nonumber \\
&+&
(-6.63694 - 1.20013 n_f - 0.00518 n_f^2 ) \, a_s^3
\nonumber \\
&+&(-156.61 + 18.77\, n_f - 0.7974\, n_f^2  \nnb\\
&&+ 0.0215\,  n_f^3 ) \, a_s^4~,
\label{eq:R_th}
\eea
where the $\alpha_s^2$, $\alpha_s^3$ and $a_s^4$ contributions have been computed respectively in \cite{DINE,LARIN,KUHN}. 
The  perturbative quantity $\delta^{(0)}$ can be evaluated using Fixed Order perturbation theory (FO)
 or using the so-called  ``Contour Improvement'' (CI). To order $\alpha_s^4$, it reads \cite{BNP0,BNP,LEDIB,PIVOV,KUHN}:
\bea
\delta^{(0)}_{FO} &=&   a_s + 5.202 \ a_s^2 +  26.366 \ a_s^3  +127.079 \, a_s^4 
\nnb,
\\
\delta^{(0)}_{CI} &=& 
 1.364 \, a_s+ 2.54 \, a_s^2 + 9.71 \, a_s^3
 + 64.29 \, a_s^4~.
\label{eq:deltath1}
\eea
which, for a reference value $\alpha_s(M_\tau)=0.34$, gives :
\beq
\delta^{(0)}_{FO} = 0.2204 ~~~{\rm and}~~~~~\delta^{(0)}_{CI} = 0.1985~,
\label{eq:deltath}
\eeq
where one can notice a slight difference in the evaluation of $\delta^{(0)}_{CI}$ from different authors \cite{DAVIER,KUHN,BJ,MALT}. This difference will only affect the last digit in the estimate of $\alpha_s$.
We use the recent experimental value: $R_{\tau,V+A} = 3.479 \pm 0.011$ \cite{DAVIER,PICH},
and the value of the sum of standard power corrections $ \delta_{\rm st}$ in Eq. (\ref{eq:pcst}), from which we deduce:
  \beq
 \delta^{(0)}\vert_{\rm st}^{\rm exp}= 0.2081~(38)_{\rm exp}~(11)_{\rm st}~.
 \label{eq:deltapheno}
 \eeq
 Equating Eqs. (\ref{eq:deltath1}) and (\ref{eq:deltath}) with Eq. (\ref{eq:deltapheno}), we obtain to order $\alpha_s^4$ within the SVZ-expansion:
\bea
\als(M_\tau)\vert_{\rm st}&=& 0.3294~ (33)_{\rm ex}(10)_{\rm st} ~({\rm FO})\nnb\\
&=& 0.3516~ (44)_{\rm ex}(13)_{\rm st}~(\rm{CI})~,
\label{eq:alpha0}
\eea
which, at the $M_Z$ scale, becomes:
\bea
\als(M_Z)\vert_{\rm st}&=& 0.1197~ (5)_{\rm ex}(2)_{\rm st}(2)_{\rm ev} ~({\rm FO})\nnb\\
&=& 0.1223~ (4)_{\rm ex}(1)_{\rm st}(2)_{\rm ev}~(\rm{CI})~.
\label{eq:alpha1}
\eea
At this stage, the theoretical errors do not take into account the estimate of the higher order terms
which we shall discuss in the next section.
One can compare these results with some other determinations from $\tau$-decay within the SVZ expansion given in Table~\ref{tab:alpha} and the runned value at $M_Z$ with the one from the $Z$-width \cite{KUHN} and from a global fit  of electroweak data at ${\cal O}(\alpha_s^4)$~\cite{DAVIER}:
\beq
 \alpha_s(M_Z)\vert_{N^3LO} =  0.1191~ (27)_{\rm exp}(1)_{\rm th}~,
\label{eq:alphamz}
\eeq
and with the most recent world average \cite{BETHKE}:
\beq
\als(M_Z)\vert_{\rm world}=0.1189~(10)~.
\label{eq:alphaworld}
\eeq
 One can notice a quite good agreement within the errors but, like the ones from \cite{DAVIER} and \cite {KUHN}, these values are on the high side of the world average. Then, one may wonders on the possible  effects on these results from some other  (often unwanted) effects beyond the SVZ expansion. This will be the subject of the next section.

{\scriptsize
\begin{table}[hbt]
\setlength{\tabcolsep}{0.45pc}
 \caption{\scriptsize Different determinations of $\alpha_s(M_\tau)$ from hadronic $\tau$-decay 
 to order $\alpha_s^4$, where the difference comes from the variant in the appreciation
 of the non-calculated higher order corrections and of the choice of the PT series. $\la~\ra$
 means average of FO and CI and $ev$ means evolution from $M_\tau$ to $M_Z$. 
 }
\scriptsize
\begin{tabular}{llll}
&\\
\hline
\hline
\\
PT&$\alpha_s(M_\tau)$&$\alpha_s(M_Z)$&Ref.\\
&\\
\hline
\hline
\\
&\\
CI&$0.3440~(50)_{\rm ex}(70)_{ \rm th}$&$0.1212 ~(5)_{\rm ex} (8)_{ \rm th}(5)_{\rm ev}$& \cite{DAVIER}\\
&\\
$\la~\ra$ &$0.3320~ (50)_{\rm ex} (150)_{ \rm th}$&$0.1202 ~(6)_{\rm ex}(18)_{ \rm th}(3)_{\rm ev}$& \cite{KUHN}\\
&\\
FO&$0.3156~(30)_{\rm ex}(51)_{\rm th}$&$0.1180~(4)_{\rm ex}(6)_{\rm th}(2)_{\rm ev}$&\cite{BJ}\\
&\\
CI&$0.3209~(46)_{\rm ex}(118)_{\rm th}$&$0.1187~(6)_{\rm ex}(15)_{\rm th}(3)_{\rm ev}$&\cite{MALT}\\
&\\
\hline
&\\
&&&This work\\
&\\
FO &$0.3294~ (33)_{\rm ex}(10)_{\rm st}$&$0.1197~ (5)_{\rm ex}(2)_{\rm st}(2)_{\rm ev}$&Eq. (\ref{eq:alpha0})\\
CI&$0.3516~ (44)_{\rm ex}(13)_{\rm st}$&$0.1223~ (4)_{\rm ex}(1)_{\rm st}(2)_{\rm ev}$& \\
&\\
FO &$0.3276~(34)_{\rm ex}(86)_{\rm th}$&$0.1195~(4)_{\rm ex}(10)_{\rm th}(2)_{\rm ev}$&Eq. (\ref{eq:alphanst})\\
CI &$0.3221~(48)_{\rm ex}(122)_{\rm th}$&$0.1188~(6)_{\rm ex}(15)_{\rm th}(2)_{\rm ev}$&\\
$\la~\ra$&$0.3249~(29)_{\rm ex}(75)_{\rm th}$&$0.1192~(4)_{\rm ex}(9)_{\rm th}(2)_{\rm ev}$&
Eq. (\ref{eq:alfaaverage})\\
&\\
\hline
&\\
Z&&$0.1191~(27)_{\rm ex}(2)_{\rm th}$&\cite{KUHN,DAVIER}\\
$\la~\ra$&&0.1189~(10)&\cite{BETHKE,PDG}\\
&\\
\hline
\hline
\end{tabular}
\label{tab:alpha}
\end{table}
}
\nin
\section{Power corrections beyond the SVZ expansion}
\nin
To the previous standard contributions of the OPE, there are also other NP contributions which have been
overlooked or/and not considered carefully in the existing literature. These contributions  are expected to be present as we work with truncated (non) perturbative QCD series which will never describe exactly inclusive processes like e.g. $\tau$-decay, $e^+e^-$, but instead can only give a smearing of it \cite{POGGIO} \footnote{As argued in \cite{SHIF}, lattice calculations may cannot help for solving this issue.}. Moreover, even if we are able to add the number of terms we want into the (non) perturbative QCD series,  we will not be able to get exactly the QCD two-point correlator involved in an inclusive process such as $\tau$-decay and $e^+e^-$ because the QCD series are factorially divergent.  
\\
{\b \bf Small size instantons} \\
Direct instantons are expected to be present in QCD for explaining the $\eta' - \pi$ mass shift (the so-called $U(1)_A$ axial problem \cite{U1}). At large $Q^2$, it will be highly suppressed as it can be parametrized by an  operator of  high-dimension $D=9$.  Its quantitative effect has been discussed in previous QSSR literature and has lead to some controversy \cite{SNB}. A phenomenological fit from $e^+e^-$ data leads to \cite{SNG}:
\bea
\delta_{\rm V,inst}&\simeq & 20\delta_{\rm V+A,inst}\nnb\\
&\approx& -(0.7\pm 2.7)10^{-3} ~.
\eea
Taking for a reference CI perturbative series and the corresponding value $\alpha_s(M_\tau)$ in Eq. (\ref{eq:alpha0}), it would induce a shift on the value of $\alpha_s$:
\beq
\delta\alpha_s(M_\tau)\vert_{\rm inst}=(0.4\pm 1.6)10^{-4}~,
\eeq
which is invisible at $M_Z$ within the present accuracy.\\
{\b \bf Quark-hadron duality violation} \\
A description of the measured spectral function which is not possible using PT QCD alone has been
initially discussed in \cite{SHIF}.  This so-called ``duality violation" (DV) (deviation of the prediction of the QCD truncated series from the measured spectral function) can be quantified as \cite{PERIS}:
\beq
{\rm Im} \Pi(t)\vert_{\rm DV}=\kappa e^{-\gamma t}\sin (\alpha+\beta t)\theta (t-t_{\rm min})~,
\eeq
where $\kappa,~\gamma,~\alpha$ and $\beta$ come from the fit of the $\tau$-data for $t_{\rm min}\simeq 1.1~{\rm GeV^2}\leq t\leq M^2_\tau$, and depend on the channels studied. DV induces an effect~\cite{PERIS}:
\beq
 \delta_{DV,V}=-(15\pm 9)10^{-3}~,~~\delta_{DV,A}\simeq (2\pm 2)10^{-4},
\eeq
respectively in the $V$ and $A$ channels, where $ \delta_{DV,V}$ (which is model dependent)  can be relatively large in agreement with the rough estimate $\vert \delta_{DV,V}\vert \approx 3\%$ 
obtained in \cite{SHIF}~\footnote{An extension of the model to fit $e^+e^-$ data above the $\tau$ mass would lead to a slightly lower value of about $-(2/3)(6.5\pm 4.2)10^{-3}$ though consistent with the former within the errors.}, while $\delta_{DV,A}$ is almost negligible and remains to be understood. DV would induce an effect in units of $10^{-4}$:
\beq
\delta\alpha_s(M_\tau)\vert_{\rm DV}=175\pm 101,~~ \delta\alpha_s(M_Z)\vert_{\rm DV}=18\pm 17.
\eeq
{\b\bf Tachyonic gluon  from $e^+e^-$ and $\pi$ sum rules}\\
This  mass can induce a new $1/Q^2$-term not present in the original SVZ expansion \footnote{An earlier phenomenology of the $1/Q^2$-term in $\tau$-decay and $e^+e^-$ has been discussed in \cite{ALT,SND2}.}. Some eventual origins of this term \cite{ZAKA,ZAK,AKHOURY,CHERNO,D2} and its phenomenological applications  \cite{SND2,CNZ,SNREV,SNZ1,SNQ,SNmse,SNmstau,RAKOW} have been discussed in the literature, as well as its relation with the short distance linear part of the QCD heavy quark potential.
Here, we shall consider that the $1/Q^2$-term is purely of short distance nature and can mainly emerge from the resummation of the infinite terms of the pQCD series (UV renormalon).  
It is expected to provide  a phenomenological parametrization of the unknown higher order terms of the PT series as an alternative to the estimates in the existing literature \cite{BNP,LEDIB,ALEPH,OPAL,DAVIER,KATAEV} and to the large $\beta_0$-approximation of the UV renormalon contribution \cite{ZAKA,ZAK,BJ}. However, its size can depend on the order at which the PT series is truncated and may (in principle) disappear if  infinite terms of the series are known \cite{RAKOW,ZAK}~\footnote{More discussions on its motivation and r\^ole  in some other hadronic channels and QCD phenomena  will be reported in a forthcoming work.}. Its contribution to ${\rm R}_\tau$ is \cite{CNZ}:
\bea
M_\tau^2\delta^{(2)}_{\rm V,tach}&=&M_\tau^2\delta^{(2)}_{\rm A,tach} =-2\times 1.05a_s\lambda^2~,
 \eea
 where $\lambda^2$ is the tachyonic gluon mass estimated from an overall fit of the pion sum rule \cite{CNZ,SNREV} and
 of the I=1 part of the $e^+e^-\rar$ hadrons data \cite{SNe,SND2,CNZ}. The average of the two determinations gives \cite{SNREV}:
 \beq
 a_s\lambda^2=-(0.07\pm 0.03)~{\rm GeV}^2~.
 \label{eq:tach}
 \eeq
 Its presence will only affect slightly the previous determinations of the condensates having higher dimensions, where its correlation to  the estimate of $\la \alpha_s G^2\ra$ from $e^+e^-$ has been studied in \cite{SNe}. 
 As originally introduced, it will be, instead, relevant to the estimate of unknown higher order PT series, which will be discussed in the next section.  It would induce, in the V+A channel, an effect :
 \beq
 \delta^{(2)}_{\rm  tach}\vert_{\rm pheno}=(46\pm 20) 10^{-3}~.
 \label{eq:corrtach}
\eeq
{\b\bf Tachyonic gluon  from large $\beta_0$-approximation}\\
It is instructive to compare the previous value of the tachyonic gluon (TACH) contribution  from the one which would be obtained from the large $\beta_0$-approximation. In so doing, we take advantage of the analysis in \cite{BJ}, 
where, for  a reference value $\alpha_s(M_\tau)=0.34$, one can predict for the Borel summed PT series: \beq
\delta^{(0)}_{\rm \beta}=0.2371~.
\eeq
Truncating the PT series known until $n=4$, one can deduce from Eqs. (\ref{eq:deltath1}) and (\ref{eq:deltath}) in units of $10^{-3}$:
\bea
 \delta^{(2)}_{\rm  tach}\vert_{\beta}\equiv  \delta^{(0)}_{\rm\beta}-\delta^{(0)}_4&\simeq&  17\pm 0.5 ~{\rm (FO)}
 \nnb\\
 &\simeq& 39\pm 5~{\rm (CI)}
 ~,
 \label{eq:tachbeta}
\eea
which agrees nicely with the phenomenological fit in Eq. (\ref{eq:corrtach}). We have estimated the error from the deviation of the large $\beta_0$-approximation from the sum of the calculated terms of the series truncated at $n=4$. However, as can been observed from \cite{DAVIER,BJ}, though the CI converges faster than FO, it will never reach the sum of the large $\beta_0$-approximate result,  because the CI series stabilizes for $n\geq 5-6$ well below the exact value, while FO slowly reaches the exact result for $n\geq 7$. Therefore, the inclusion of  TACH with the value in Eq. (\ref{eq:tachbeta}) is necessary for the PT series to reach the ``exact result". In the case of FO, the TACH contribution will decrease for increasing numbers of term in the PT series if one follows the analysis in \cite{BJ}, while for CI, it will remain almost constant. 
For definiteness, we shall use the large $\beta_0$-approximate result in  Eq. (\ref{eq:tachbeta}), which can be more appropriate and accurate for this specific channel than the phenomenological fit.  For CI, it would induce an effect in units of $10^{-4}$:
\beq
\delta\alpha_s(M_\tau)\vert_{\rm tach}\simeq -491\pm 63~,~~ \delta\alpha_s(M_Z)\vert_{\rm tach}\simeq-61\pm 8~.
\eeq
\b {\bf Total contributions} \\
Adding the previous contributions beyond the SVZ expansion, we obtain an estimate of the total ``new " contributions in the V+A channel (in units of $10^{-3}$):
\bea
\delta_{\rm nst}\equiv \delta_{\rm inst}+\delta_{DV}+\delta^{(2)}_{\rm tach}&=& (2.0\pm 9.4)~{\rm FO}\nnb\\
&=& (24.0\pm 10.6)~{\rm CI}~,
\label{eq:pcnst}
\eea
where there is a partial cancellation between the DV and TACH contributions.  
These effects can be about a factor 3 larger than the standard SVZ non-perturbative contributions given in Table~\ref{tab:svz} but remain relatively small compared to the PT contributions of about 20\% which we shall discuss in the next section.  
\section{  $\delta_{\rm st+nst}$ corrections to $\alpha_s(M_\tau)$ }
\nin
Adding the sum of $\delta_{\rm nst}$ corrections in Eq. (\ref{eq:pcnst}) to 
the value of  $\delta_{\rm st}$  in Eq. (\ref{eq:pcst}), the 
estimates in Eqs. (\ref{eq:deltapheno}) to (\ref{eq:alpha1}) become to ${\cal O}(\alpha_s^4)$:
\bea
 \delta^{(0)}_{\rm st+nst}\vert_{\rm exp}&=& 0.2061~(38)_{\rm ex}(11)_{\rm st}(94)_{\rm nst} ~{\rm (FO)}\nnb\\
 &=& 0.1841~(38)_{\rm ex}(11)_{\rm st}(94)_{\rm nst} ~{\rm (CI)}~.
 \eea
 implying:
\bea
\als(M_\tau)&=& 0.3276~(34)_{\rm ex}(10)_{\rm st}(85)_{\rm nst} ~({\rm FO})\nnb\\
&=& 0.3221~(48)_{\rm ex}(14)_{\rm st}(121)_{\rm nst}~({\rm CI})~.
\label{eq:alphanst}
\eea
However, as explained in the previous section, the result from CI  is preferred than from FO
due to the faster convergence of the PT series, where the difference between the Borel summed
series and the known PT series  is expected to be better reproduced by $ \delta^{(2)}_{\rm tach}$
given in Eq.~(\ref{eq:corrtach}) which remains (almost) constant \cite{BJ}, when higher order terms of the series will be available in the future. 

\section{Comparison of $\alpha_s$ from  $\tau$-decay and $Z$-data.}
\nin
 We run  our previous result in Eq. (\ref{eq:alphanst}) at $M_Z$ and obtains:
\bea
\als(M_Z)\vert_\tau&=& 0.1195~(4)_{\rm ex}(1)_{\rm st}(10)_{\rm nst}(2)_{\rm ev} ~({\rm FO})\nnb\\
&=&0.1188~(6)_{\rm ex}(2)_{\rm st}(15)_{\rm nst}(2)_{\rm ev} ~({\rm CI})~.
\label{eq:alphastef}
\eea
Both results agree quite well each other and with the one from $Z$-data in Eq. (\ref{eq:alphamz}) and from the world average in Eq. (\ref{eq:alphaworld}). They indicate that the presence of the TACH reduces the difference between the FO and CI results by a factor 4 [comparison of Eqs. (\ref{eq:alpha0}), (\ref{eq:alpha1}) and (\ref{eq:alphastef})], which is not the case of the ones in the existing literature. 
We consider, as a final result, the average of the two determinations FO and CI 
to ${\cal O}\ga\alpha_s^4\dr$ and in the $\overline{MS}$-scheme:
\bea
\la\als(M_\tau)\ra&=& 0.3249~(29)_{\rm ex}(9)_{\rm st}(74)_{\rm nst} \lrar\nnb\\
\la\als(M_Z)\ra\vert_\tau&=&0.1192~(4)_{\rm ex}(1)_{\rm st}(9)_{\rm nst}(2)_{\rm ev}~. 
\label{eq:alfaaverage}
\eea
to which correspond:
\bea
\Lambda_3&=&353~(6)_{\rm ex}(2)_{\rm st}(14)_{\rm nst}~{\rm MeV}~~\lrar\nnb\\
\Lambda_4&=&307~(6)_{\rm ex}(2)_{\rm st}(14)_{\rm nst}(2)_{\rm ev}~{\rm MeV}~,\nnb\\
\Lambda_5&=& 223~(5)_{\rm ex}(1)_{\rm st}(11)_{\rm nst}(1)_{\rm ev}~{\rm MeV}~.
\eea
In Table \ref{tab:alpha}, we  compare our results  
with the ones obtained using models which differ in the estimate of the strength of the unknown higher order PT terms: \\
\b  The most popular \cite{BNP,LEDIB,ALEPH,OPAL,DAVIER} is the estimate of the coefficient of the next order term either as equal to the last term of the PT series or a prediction based on a geometric growth of the PT coefficient \cite{BNP} or using \cite{KATAEV} the principles of ``Fastest Apparent Convergence "(FAC) \cite{GRUN} of ``Minimal Sensitivity" (PMS) \cite{STEVE} \footnote{Some alternative approaches can be found in \cite{ALTERN}.} . In \cite{DAVIER} CI is preferred over FO as the series converge faster. \\
\b Using an analogous estimate of the higher order terms, Ref. \cite{KUHN} takes the average of the FO and CI results,  where 1/3 of the theoretical error (-0.005) comes from the estimate of the unknown $\alpha_s^5$ term. \\
 \b An alternative approach is the use of a toy model for the Borel transform of the Adler function beyond the $n=4$ order. This model favours FO over the CI   \cite{BJ}, where the main theoretical error comes from the renormalisation scale dependence of the FO.\\
 \b In \cite{MALT}, CI has been used in connection with Finite Energy Sum Rules having some specific weights.\\
\b In the present  work, all possible (standard $\delta_{\rm st}$ and non-standard $\delta_{\rm nst}$) QCD power corrections are considered in the extraction of $\alpha_s(M_\tau)$. One can notice, from Table \ref{tab:alpha}, a good agreeement between different determinations, despite various appreciations on the estimate of higher order terms. However, according to the analysis in Ref. \cite{PERIS}, none of the previous estimates have carefully considered the quark-hadron duality violation (DV) effects which cannot be neglected. The DV effects shift the results of \cite{DAVIER} and \cite{KUHN} to the high-side of the world average.  In this paper, we point out, that the tachyonic gluon mass partially cancels the DV effects  and reduces the usual systematic difference between the FO and CI results by a factor 4 [comparison of Eqs. (\ref{eq:alpha0}), (\ref{eq:alpha1}) and (\ref{eq:alphastef})].  \\
\b There is a remarkable agreement between the two determinations from $\tau$ [Eqs. (\ref{eq:alphastef}) and  (\ref{eq:alfaaverage})] and $Z$ [Eq. (\ref{eq:alphamz})] decays
obtained  at the same $N^3LO~{\cal O}(\alpha_s^4)$ level of accuracy and at different regions of energy which demonstrates the running strong coupling $\alpha_s$ of QCD predicted by asymptotic freedom. At present, the two values of $\alpha_s(M_Z)$ from $\tau$ and $Z$-decays are the most accurate determinations available today compared to the other determinations compiled in  \cite{PDG,BETHKE}. Though the non-standard power corrections give the largest errors of $\alpha_s(M_\tau)$, its accuracy is not largely affected by these new contributions.  
\section{Tachyonic gluon mass to $\vert V_{us}\vert$  and $m_s$}
\nin
 \b The CKM angle { $\bf\vert V_{us}\vert$}\\
  It can be determined  accurately \cite{PICH}:
\beq
\vert V_{us}\vert=0.2212~(31)_{\rm exp}(5)_{\rm th}~,
\eeq
 from the observable:
 \beq
\vert V_{us}\vert ^2= {R_{\tau,S}\over {\ga R_\tau/ \vert V_{ud}\vert^2\dr}-\delta R_{\tau}}~,
\eeq
 if one uses the experimental values:
 \bea 
 {R_\tau\over \vert V_{ud}\vert^2} = 3.6661~(12)~,~~{R_{\tau,S}\over \vert V_{us}\vert^2}=0.1686~(47)~,
 \eea
  while the theoretical input only enters into the SU(3) breaking quantity:
  \beq
  \delta R_\tau=0.227(54)~,
  \eeq
   which is one order of magnitude smaller than $R_\tau$. About 87\% of the error in $\delta R_\tau$ is mainly due to the value of $m_s$, which has been taken in \cite{PICHms} to be: $m_s(2~{\rm GeV})=(96\pm 10)$ MeV. This error  reduces further by a factor 2 if one uses the average quoted in Eq. (\ref{eq:msaverage}). 
One may wonder on the TACH contribution to $ \delta R_\tau$. TACH effects to the SU(3) breaking part of the hadronic correlator are \cite{CNZ}:
\bea
\delta\Pi(Q^2)_{V+A}^{(1)}&=&{1\over 16\pi^2}{a_s\lambda^2\bar m^2_s\over Q^4}\ga -{100\over 3}+16\zeta(3)\dr\nnb\\
\delta\Pi(Q^2)_{V+A}^{(0)}&=&{3\over 8\pi^2}{a_s\lambda^2\bar m^2_s\over Q^4}~,
\eea
where $\zeta(3)=1.202056...$ is the Riemann zeta function.
Its contribution to $\delta R_\tau$ vanishes to leading order  in $\alpha_s$ like other dimension $D=4$ operators and can be neglected in the determination of $\vert V_{us}\vert$ within the present accuracy of $\delta R_\tau$.  \\
\b {\bf  $\bf \bar m_s$ from $\tau$-decay and $e^+e^-$}\\
It has been determined from the SU(3)-breaking $\tau$-decay widths \cite{PICH,PICHms}:
 \beq
 \delta R_\tau\equiv  {R_\tau \over \vert V_{ud}\vert^2}-{R_{\tau,s} \over \vert V_{us}\vert^2}~,
 \label{eq:mstau}
 \eeq
 or from $\tau$-like  decays in $e^+e^-$ using the difference between the isovector and $\phi$-meson channels \cite{SNmse,SNQ}:
 \beq
 \Delta_{1\phi}\equiv R_{\tau,1}-R_{\tau,\phi}~,
 \eeq
 or with the difference of the usual exponential inverse Laplace or/and FESR sum rules \cite{SNmstau,SNQ}. \\
However, TACH contribution is larger in the case of LSR and FESR used in \cite{SNmstau,SNQ} for determining $m_s$, because in these observables, there is no more a cancellation of the $D=4$ contribution occured in $R_\tau$. A correlation between the values of $m_s$ and $a_s\lambda^2$ has been studied. FESR analysis with the value of $a_s\lambda^2$ in Eq. (\ref{eq:tach}) gives to order $\alpha_s^3$:
\bea
\bar m_s(2~{\rm GeV})&=& 100~(34)~{\rm MeV}~:~~~~~{\rm e^+e^-}~,\nnb\\
&=& 93~(30)~{\rm MeV}~:~~~~ \tau{\rm-decay}~,
\eea
in good agreement with the weighted average in Eq. (\ref{eq:msaverage}),  but disfavours the value $\lambda=0$ to which would correspond a too small $m_s$ mass but having larger errors. \\
\b {\bf  $\bf \bar m_s$ from the (pseudo)scalar channels} \\
One should also note that TACH contribution in the pseudoscalar pion and kaon sum rule channels reduces the estimate obtained for $\lambda=0$ to about 5\% \cite{CNZ}:
 \bea
 (\bar m_u+\bar m_d)(2~{\rm GeV})&=& 8.6~(2.1 )~{\rm MeV}~,~~\lrar\nnb\\
 \bar m_s(2~{\rm GeV})&=& 105~(26 )~{\rm MeV}~,
\eea
which is, in better agreement with the average in Eq. (\ref{eq:msaverage}). In the same time, it improves the duality between the experimental and theoretical side of the sum rules.

\section{Summary and conclusions}
\nin
We have reexamined the determinations of the SVZ power corrections and their effects on the extraction of $\alpha_s$ from $\tau$-decay data :\\
\b $\la \alpha_s G^2\ra$ is better determined from QSSR \cite{SNB,SNe,SNG} than from $\tau$-decay with the robust estimate  in Eq. (\ref{eq:e+e-}).\\
\b In the V channel, $\tau$-decay and $e^+e^-$ sum rules lead to a value of the four--quark condensate 
$\rho\alpha_s\la\bar uu\ra^2$ in Eq. (\ref{eq:4quark}) which violates the vacuum saturation estimate by a factor 3 [Eq. (\ref{eq:rhofactor})] and confirms previous QSSR results from different hadronic channels
\cite{SNB,SNe,TARRACH,PEROTTET,SNZ1,RAF04,SNVA,DOSCH}. \\
\b Using the V+A $\tau$-decay channel, which is the most sensible one for extracting $\alpha_s$, we show in Eqs. (\ref{eq:alphastef}) and (\ref{eq:alfaaverage}), our estimate to ${\cal O}\ga\alpha_s^4\dr$ which takes into account the effects of power corrections beyond the standard SVZ expansion. Instead of ruining the existing estimates without these non standard effects, these new contributions  improve the agreement of the $\alpha_s$ value in Eqs. (\ref{eq:alphastef}) and  (\ref{eq:alfaaverage}) with the one in Eq. (\ref{eq:alphamz}) from the direct determinations at the $Z$-mass from the $Z$-width and global fit of electroweak data as well as from the recent world average in Eq. (\ref{eq:alphaworld}). The 
determinations of $\alpha_s(M_\tau)$ from $\tau$-decay lead to the most accurate value of $\alpha_s(M_Z)$ available today and demonstrates with high accuracy the running of the QCD coupling as expected from asymptotic freedom !\\
\b Finally, we have discussed the r\^ole of the tachyonic gluon mass by taking the example of the determinations of $\vert V_{us}\vert$ from $\tau$-decay and of $m_s$ from $\tau$-decay, $e^+e^-$ and pseudoscalar channels. Though its effect is negligible in the $\tau$-like decay sum rules due to the vanishing of the $D=4$ contribution, to leading order, in this particular observable, it plays an important r\^ole in the LSR and FESR approaches. 
\section*{Acknowledgement} 
\nin
It is a pleasure to thank Santi Peris and Valya Zakharov for communications and for reading the manuscript.


\end{document}